\documentclass{article}
\usepackage[numbers]{natbib}
\usepackage{amssymb}
\usepackage{amsmath}
\usepackage{graphicx}  
\usepackage{hyperref}

\usepackage{color}

\usepackage{soul}

\usepackage[normalem]{ulem}

\usepackage[utf8]{inputenc}

\usepackage{tikz}

\usepackage{amssymb}
\usepackage{amsmath}
\usepackage{graphicx}  
\usepackage{hyperref}

\usepackage{color}

\usepackage{soul}

\usepackage{authblk}
\title{Signatures in the local interstellar medium of electromagnetic radiation produced by neutron star gravitational waves}
\author[1,3]{Preston Jones}
\author[1]{Pragati Pradhan}
\author[2]{Douglas Singleton}
\affil[1]{Embry-Riddle Aeronautical University,
3700 Willow Creek Road,
Prescott, AZ 86301, USA}
\affil[2]{California State University, Fresno, 5241 N Maple Ave, Fresno, CA 93740, USA}
\affil[3]{Corresponding author email: jonesp13@erau.edu}

\begin{document}




\date{\today}

\maketitle

\begin{abstract}
The Voyager spacecraft have been detecting plasma oscillations of a few kHz for over 40 years. However, there is still no widely accepted hypothesis for the astrophysical source of the oscillations. Having entered the local interstellar medium, these detections have persisted, adding to the challenge of identifying the astrophysical source. Neutron stars were one of the earliest hypothesized candidates of the electromagnetic radiation responsible for the detected plasma oscillations, but no theoretical mechanism could be identified to explain the frequency of the oscillations. We explore the recently developed theory of co-production of electromagnetic radiation in gravitational wave backgrounds as a possible source of these low-frequency oscillations and investigate whether neutron stars could be the astrophysical source.
\end{abstract}






\section{Introduction}

The Voyager spacecraft's scientific mission has been in operation for nearly half a century, and upon entering the local interstellar medium (LISM), it became the first instrument to make in situ measurements outside the heliopause. Our interest is in Voyager detections of plasma oscillation events (POE) in the LISM and ongoing efforts to determine the astrophysical source \cite{Kurth_2023} of the POE. The Voyager spacecraft have detected plasma waves in the $ \rm{kHz}$ range since at least the mid-1980s \cite{Kurth1984}. A subsequent report of later observations in 2003 \cite{https://doi.org/10.1029/2003GL018514} described the current proposed mechanism as ``generated when a strong interplanetary shock, produced by a period of intense solar activity, interacts with the heliopause''. However, there is still no meaningful agreement on the source for the POE detected in the LISM. Recently, serious doubts have been raised that either solar cycles or shocks could provide a ``definitive'' explanation for the Voyager observations \cite{Kurth_2023}. In the 1984 paper \cite{Kurth1984}, the possibility is considered that the radio emissions were generated outside the heliopause and that a fast pulsar was the astrophysical source. They ruled out a pulsar source at the time and observed, ``It is interesting to think of what other sources of radio noise lie beyond the Solar system, but perhaps fruitless to speculate on them at this time". The recent development of the theory of co-production \cite{doi:10.1142/S0218271815440174,PhysRevD.96.124030,10.1142/S0218271818470211} and identification of potential astrophysical signatures \cite{2025ApJ...989...18J} makes it possible to identify co-production from neutron stars (NS) as a potential source beyond the Solar system. This paper follows our recent work on the production of low-frequency electromagnetic radiation from gravitational wave backgrounds (co-production) \cite{2025ApJ...989...18J} and the identification of Voyager detections of POE as potential astrophysical signatures of this production. The relation of gravitational waves from NS sources and electromagnetism is similar and complementary to recent work on gravitational waves and NS magnetospheres \cite{2024MNRAS.52711198C}.

Ocker et al. \cite{2021NatAs...5..761O} published a very interesting report on the plasma oscillation events (POE) detected by Voyager after traveling beyond the heliopause. They describe weak narrow-band plasma waves of approximately  $3 ~ \rm{kHz}$ in interstellar space. The potential mechanism for the production of the oscillations is primarily attributed to processes occurring within the heliosphere or near the heliopause. We would like to bring attention to another possible mechanism associated with co-production and, in particular, the electromagnetic radiation production described in \cite{PhysRevD.96.124030,10.1142/S0218271818470211} for gravitational waves (GW) from NS. Considering the frequencies of GW vacuum production of electromagnetic radiation or co-production, the association of the Voyager narrow-band observations with NS sources is a possible explanation (see Section \ref{NSsource}). The scenario for Voyager detections of POE as signatures of co-production is: NS production of GW at twice the NS rotation frequency, co-production of electromagnetic radiation at twice the GW frequency, and co-production induced POE in the LISM. The two doublings of the NS spin frequency are consistent with co-production as one possible source of the observed POE for the most energetic NS with, e.g., spin frequencies of $\sim700~\text{Hz}$, consistent with $\sim 3 ~ \text{kHz}$ POE frequencies observed by Voyager. With a single doubling for GW production, the search window for coincident GW should be nominally $1.5 ~ \text{kHz}$.

The association of the more energetic POE with a NS source is complicated by the observed frequency changes and persistence of the plasma oscillations. Frequency drift and persistence would not generally be expected from co-production driven by energetic GW bursts from NS glitches and anti-glitches. Energetic or long-duration GW emission from NS is possible \cite{2020MNRAS.498.3138Y,2022PhRvD.105b2002A}, although other
explanations of the observed persistence may be more plausible. An interesting possibility would be NS glitches from accretion processes discussed in Section \ref{NSsources}. It is noteworthy that accretion processes would be associated with the cosmic rays \cite{1997ApJ...484..323V,Cameron1965,10.1093/mnras/275.1.115,Rocamora_2024} which were also detected by Voyager \cite{2021NatAs...5..761O,2021AJ....161...11G,2025ApJ...993...81C}, adding support to the possibility that NS are the source of POE detected in the LISM. It is also possible and perhaps more likely that the persistence and the frequency changes are a natural property of plasma oscillations in the LISM. If the astrophysical origin of co-production, stimulating POE in the LISM, is a NS then the NS of Cassiopeia A and Vela Jr. are likely candidates. Both Cassiopeia A and Vela Jr. \cite{2025ApJ...986..202M} are expected to produce energetic and possibly detectable GW, which should include co-production of low-frequency electromagnetic radiation associated with GW bursts from NS glitches.

\section{Co-production intensities from neutron star sources} \label{NSsource}

The conversion of gravitons to photons at tree level, $g + g \rightarrow \gamma + \gamma$, has been studied in \cite{Skobelev1975}, but the cross-sections are far too small to be observable. Semi-classical conversion is also possible as demonstrated by solving the covariant Maxwell equations in the background of GW \cite{doi:10.1142/S0218271815440174,10.1142/S0218271818470211,2025ApJ...989...18J,PhysRevD.95.065010}. Since the semi-classical production of radio emissions is at twice the frequency of the GW, NS are a potential source of the signals detected by Voyager beyond the Solar System. The maximum observed neutron-star rotation rate is approximately $700~\mathrm{Hz}$, as in PSR J1748--2446ad. Rotation-dependent gravitational wave production would be at twice the rotation frequency \cite{2020MNRAS.498.3138Y,2022PhRvD.105b2002A}. This, along with the vacuum production of radio waves at twice the GW frequency \cite{PhysRevD.95.065010}, leads to two frequency doublings in the processes from NS rotation to plasma oscillations. This makes a subset of energetic neutron stars plausible candidate sources for the extrasolar radio emission observed by Voyager.

We first demonstrate the potential of co-production from NS gravitational waves, as the astrophysical source of the POE detected by Voyager, by calculating the intensity of the low-frequency electromagnetic radiation at the source and relating this to the distance from the source. 
The Lagrange density for an electromagnetic field in the Lorenz gauge is
\[
 {\cal L}_{\rm em}
 =-\frac{1}{2}\partial_{\mu}A_{\nu}\partial^{\mu}A^{\nu},
\]
where
\[
 A_{\mu}(k,\lambda,x)
 =\epsilon_{\mu}^{(\lambda)}
  \varphi^{(\lambda)}(k,x).
\]
Here $x^{\mu}=(t,x,y,z)$ is the spacetime position, $k^\mu$ is the four-wavevector characterizing the plane-wave mode, the Greek spacetime indices $\mu,\nu=0,1,2,3$, and we are using the signature $(-,+,+,+)$.  The label $\lambda=1,2$ denotes the two physical electromagnetic polarizations, $\epsilon_{\mu}^{(\lambda)}$ is the corresponding polarization four-vector, and $\varphi^{(\lambda)}(k,x)$ is the scalar mode amplitude. For the plane wave propagating in the $+z$ direction used below, $k\equiv |\boldsymbol{k}|$ denotes the wave number.  In units with $c=1$, the dispersion relation is $\omega=k$, so that the phase can be written as
\[
 k(z-t)=kz-\omega t.
\]
Thus, $t$ is the time coordinate and is independent of $k$.

For a single polarization of gravitational radiation the field equations for the vector potential simplify to the Klein-Gordon equation with a space-time background metric $g^{\mu \nu}$,

\begin{equation}
\frac{1}{{\sqrt{-g}}} \partial_{\mu} \sqrt{-g} g^{\mu \nu} \partial_{\nu}\varphi = 0.
\label{eomvarphiA}
\end{equation}

\noindent Assuming a small amplitude, $h \ll 1$, gravitational wave background and following our previous paper \cite{2025ApJ...989...18J} the equation of motion \eqref{eomvarphiA} has the solution to fourth order in $h$,

\begin{equation}
\varphi \left( {t,z} \right) =  \frac{1}{2}h^2 e^{2ik(z - t)}  + \frac{3}{8}h^4 e^{4ik(z - t)} . 
\label{OutState}
\end{equation}

\noindent We now turn to the intensity relation between the gravitational wave background and the electromagnetic radiation production. This relation was developed in the Newman-Penrose formalism \cite{PhysRevD.96.124030,PhysRevD.95.065010,1973ApJ...185..635T} showing that the intensity ratio is $\frac{d E_{em}/dt}{d E_{gw}/dt} = \frac{{\left( {\frac{1}{{4\pi }}\left| {\Phi _2 } \right|^2 } \right)}}{{\left( {\frac{1}{{16\pi k^2 }}\left| {\Psi _4 } \right|^2 } \right)}} $, 
where $\Phi_{2}$ and $\Psi_{4}$ are Newman--Penrose scalars. The numerical subscripts label particular projections of the electromagnetic field tensor and Weyl tensor onto the Newman--Penrose null tetrad; they are not spacetime tensor indices.  In the tetrad convention used here, $\Phi_{2}$ characterizes the outgoing electromagnetic radiation and $\Psi_{4}$ characterizes the outgoing gravitational radiation.
The intensity ratio is shown to be $\frac{{\dot E_{em} }}{{\dot E_{gw} }} = 2h^2$ (assuming the weak field limit of a plane wave with a single polarization) and the production of electromagnetic radiation by gravitational waves is on the order of $h^2$ \cite{PhysRevD.96.124030},

\begin{equation}
F_{em}   = 2 h^2 F_{gw} ~,
\label{LuminosityEMlum}
\end{equation}

\noindent where $F_{em}$ and $F_{gw}$ are the electromagnetic and gravitational flux.

In order to estimate the electromagnetic radiation production near a NS we need to establish the production ratio in terms of luminosities. That is to calculate the electromagnetic intensity from co-production of electromagnetic radiation near the neutron star source in terms of total gravitational wave luminosity, $\mathcal{L}$. The gravitational wave flux is $F_{gw}  = \frac{{c^3 }}{{16\pi G}}\left| {\dot \varepsilon } \right|^2 $ with $\left| {\dot \varepsilon } \right|^2 = h^2 \omega^2$ for small amplitude, monochromatic gravitational wave background  so that the gravitational wave intensity is given approximately by \cite{1996CQGra..13A.219S},

\begin{equation}
F_{gw}  \left( r_{0} \right)  = \frac{{c^3 }}{{16\pi G}}\left| {\dot \varepsilon} \right|^2  = \left( {3 \times 10^{35}~\rm{ \frac{{Ws^2 }}{{m^2 }}}} \right) h^2 f^2 ~,
\label{gwFlux_hlum}
\end{equation}

\noindent where $r_{0}$ is the distance of the co-production from the NS source. In the last step, we have used $\omega = 2 \pi f$ and absorbed all constant factors into a constant prefactor. Noting that the gravitational wave intensity is proportional to the luminosity at $r_{0}$, via the equation

\begin{equation}
F_{gw} \left( r_{0} \right)  = \frac{ \mathcal{L} }{4 \pi r_{0}^{2}} ~,
\label{gwFlux_hlumSimple}
\end{equation}
\noindent we can solve for the square strain amplitude,

\begin{equation}
 h^2 = \left( {3 \times 10^{-36}~\rm{ \frac{{m^2 }}{{ Ws^2}}}} \right) \frac{ \mathcal{L} }{4 \pi r_{0}^{2} f^2}    ~.
\label{strainlum}
\end{equation}

\noindent Substituting the square strain amplitude and the gravitational wave intensity into \eqref{LuminosityEMlum} gives the intensity of the co-produced electromagnetic radiation at $r_0$

\begin{equation}
F_{em} (r_0) =  \left(3 \times 10^{-36}~ \frac{\rm{ m^2 }} {\rm{ W s^2 }}  \right) \frac{\mathcal{L}^2}{8 \pi ^2 r_0 ^4 f^{2}}  ~.
\label{LuminosityEMlum4a}
\end{equation}

\noindent To turn this intensity in \eqref{LuminosityEMlum4a} into intensity at $r$, the distance from the source, we multiply by $\frac{r_0 ^2}{r^2}$ and combining constants we find

\begin{equation}
F_{em} (r) =  \left(4 \times 10^{-38}~ \frac{\rm{ m^2 }} {\rm{ W s^2 }}  \right) \frac{\mathcal{L}^2}{ r_0 ^2 r^2 f^{2}}  ~.
\label{LuminosityEMlum4b}
\end{equation}

\noindent Taking the frequency $f=1,500 ~ {\rm Hz}$, which is an approximate frequency for gravitational waves emitted by neutron stars and assuming the maximum co-production possible at $r_0 = 10^4 ~ {\rm m}$,  as an order-of-magnitude emission radius just outside the stellar surface 

\begin{equation}
F_{em} (r) =  \left(2 \times 10^{-52}~ \frac{\rm{1}} {\rm{ W}}  \right) \frac{\mathcal{L}^2}{r^{2}}  ~.
\label{LuminosityEMlum4}
\end{equation}

We can now compare the intensity of the co-production at a distance $r$ from the source to the measured intensity of the POE detections by Voyager. The detected intensities can be calculated from the measured E-field strengths on the order of $10^{-6} ~ \rm{\frac{V}{m}}$ to  $10^{-5} ~ \rm{\frac{V}{m}}$ \cite{Kurth_2023,doi:10.1126/science.1117425}. The minimum intensity, $E=10^{-6}~ \rm{\frac{V}{m}}$, is approximately $ I_{min}  \approx  \frac{1}{2} \epsilon_0 c E^2 \approx 10^{-15} \rm{\frac{W}{m^2}} $.
As an illustrative upper-end estimate, we take a Vela-like glitch \cite{Ball26} with an available glitch energy of order $10^{43}\ {\rm erg}$ and adopt a characteristic short-duration GW-emission timescale of order $0.1\ {\rm s}$.
The maximum possible distance for this source, consistent with co-production luminosities and assuming unit conversion efficiency is,

\begin{equation}
 r_{max}= \left[ \left(2 \times 10^{-85}~ \frac{\rm{ pc^2 }} {\rm{ W m^2 }}  \right) \frac{\mathcal{L}^2}{F_{em}} \right]^\frac{1}{2} \sim 150 ~ \rm{pc} ~,
\label{MaxD}
\end{equation}

\noindent where $F_{em} (r) = I_{min} = 10^{-15} \rm{\frac{W}{m^2}} $ and $\mathcal{L} = 10^{37} ~ \rm{W}$. 
A pulsar glitch does not necessarily produce gravitational radiation at an observable level.  A glitch is an abrupt change in the stellar rotation rate, whereas gravitational radiation requires a time-dependent, non-axisymmetric mass or current quadrupole. Proposed post-glitch mechanisms include the excitation and ringdown of fluid or crustal modes, the formation of a transient non-axisymmetric ``mountain,'' and non-axisymmetric superfluid motion during the post-glitch recovery \cite{2020MNRAS.498.3138Y,Ball26,2024CQGra..41d3001G}. The amplitude, duration, and efficiency of the resulting GW emission are strongly model dependent. We therefore use the Vela-like glitch only as a representative upper-end energetic example; we do not assume that such GW emission has been observed or that the full glitch-energy estimate is converted into gravitational radiation. 

The Vela pulsar is approximately twice this estimated limiting distance from Earth, which suggests that Vela is a less likely source for the Voyager observations. However, this order-of-magnitude calculation demonstrates that the field strength of the Voyager POE detections is consistent with the expected co-production luminosities. The most promising candidate NS sources of co-production, detected by Voyager as POE, would be the nearest ($\sim$ 100 pc) and the most energetic. Returning to \eqref{LuminosityEMlum4} and multiplying by the area, the co-production luminosity would be on the order of $10^{23} ~ \rm{W}$ and a small fraction of the GW luminosity.

\section{Neutron star sources and plasma oscillation detections} \label{NSsources}

Two characteristics of the observed POE frequencies complicate the identification of NS as the source of the radio emissions detected by Voyager: the persistence of the POE and the frequency change over time. If the associated frequency change in the POE \cite{2021NatAs...5..761O} is due to a single NS source, understanding the mechanism is not as simple as co-production driven, for example, by neutron-star quakes and associated GW bursts. 
One possible explanation for the long-term frequency evolution is co-production associated with an accreting NS, including changes in its spin or occasional accretion-associated glitches
\cite{2015A&A...578A..52D,2017MNRAS.471.4982S}. The much longer persistence of the resulting POE need not reflect the duration of the GW source and may instead be controlled primarily by the response of the rarefied LISM plasma.
Attributing the POE persistence to the LISM is then consistent with short duration glitches as the source of GW production.

As an NS accretes matter from its companion, it can experience changes in its rotational period; the magnitude and direction of the spin period change can vary greatly depending on the specific details of the binary system and the mass transfer process (see \cite{2010ApJ...722..909P} for a recent review). A 25 \% change over 7 years would require significant and potentially complex interactions between the NS and its companion, including efficient transfer of angular momentum. There are several likely scenarios, for example, during the common envelope phase, when the more massive companion star expands and engulfs the NS within its envelope \cite{1999A&A...350..928T}. A further possibility is the ongoing “recycling” for a NS in a binary system undergoing accretion from its companion, leading to a rapid increase in its spin rate. Other potential explanations are the contribution of a magnetic field to the change in the spin period of the NS or the quasi-periodic oscillations of Comptonizing regions embedded in Keplerian discs \cite{2009MNRAS.394.1463C}. 

It is also useful to compare the electromagnetic luminosity required by the present estimate with the energy available from accretion. For a neutron star of mass $M_{\rm NS}=1.4M_{\odot}$ and radius $R_{\rm NS}=10~{\rm km}$, the accretion luminosity is approximately
\begin{equation}
 L_{\rm acc}
 \simeq \eta_{\rm acc}
 \frac{G M_{\rm NS}\dot M}{R_{\rm NS}},
\end{equation}
where $\eta_{\rm acc}$ denotes the net efficiency connecting the available accretion power to the electromagnetic luminosity relevant to the present estimate. It may therefore include the efficiency of accretion-driven spin or glitch excitation, GW production, semiclassical co-production, propagation, and coupling to the LISM plasma. Using the order-of-magnitude electromagnetic luminosity $L_{\rm em}\sim10^{23}~{\rm W}$ obtained in Section~\ref{NSsource} gives
\begin{equation}
 \dot M \simeq 5\times10^{9}\eta_{\rm acc}^{-1}~{\rm g\,s^{-1}}
 \simeq  8\times10^{-17}\eta_{\rm acc}^{-1} M_{\odot}\,{\rm yr^{-1}}.
\end{equation}
This is below the canonical neutron-star Eddington accretion rate, $\dot M_{\rm Edd}\sim10^{-8}M_{\odot}\,{\rm yr^{-1}}$, unless the net efficiency is extremely small, $\eta_{\rm acc}\lesssim10^{-8}$. This estimate shows that the observed radio power does not by itself require a super-Eddington accretion rate.  It should not, however, be interpreted as a unique prediction for the binary accretion rate. In the scenario considered here, accretion may alter the stellar spin or trigger a glitch, while the GW and co-produced electromagnetic radiation draw upon the rotational and mechanical energy of the neutron star.  A quantitative mapping between $\dot M$, glitch properties, GW efficiency, and the resulting radio luminosity requires a source-specific accretion and glitch model.

The structure of the LISM would also contribute to the POE frequency due to changes in the local plasma cutoff frequency \cite{2021NatAs...5..761O,2021AJ....161...11G,2025ApJ...993...81C}. A co-production radio wave traveling through the interstellar medium would produce plasma oscillations continuously, assuming an absorption coefficient much less than one (see Section \ref{NSsource}), and the persistence would in part be a consequence of the local properties of the interstellar medium as well as the NS source.

The work of identifying the source of the low-frequency signals and cosmic rays detected by Voyager is very much an effort in multi-messenger astronomy. Previous studies have already identified an association between Voyager and ACE/CRIS cosmic-ray detections \cite{2018ApJ...858...61B,2021ApJ...913....5B}. If the Voyager POE observations are a consequence of the co-production of low-frequency electromagnetic radiation, that would help guide the search for GW from NS. The search for continuous GW from NS sources is already an active area of research \cite{2025ApJ...986..202M,2023LRR....26....3R}, and the central collapsed objects of Cassiopeia A and Vela Jr. are considered likely candidates. We would expect sufficiently energetic co-production to be observed above the noise in the LISM, which would be produced by, e.g., NS glitches, which is also consistent with the transient observations of POE by Voyager. We discuss the multi-messenger search for GW from NS in greater detail in Section \ref{NS GW detection}.


\begin{figure}[htp!]
 \centering
\includegraphics[width=115mm]{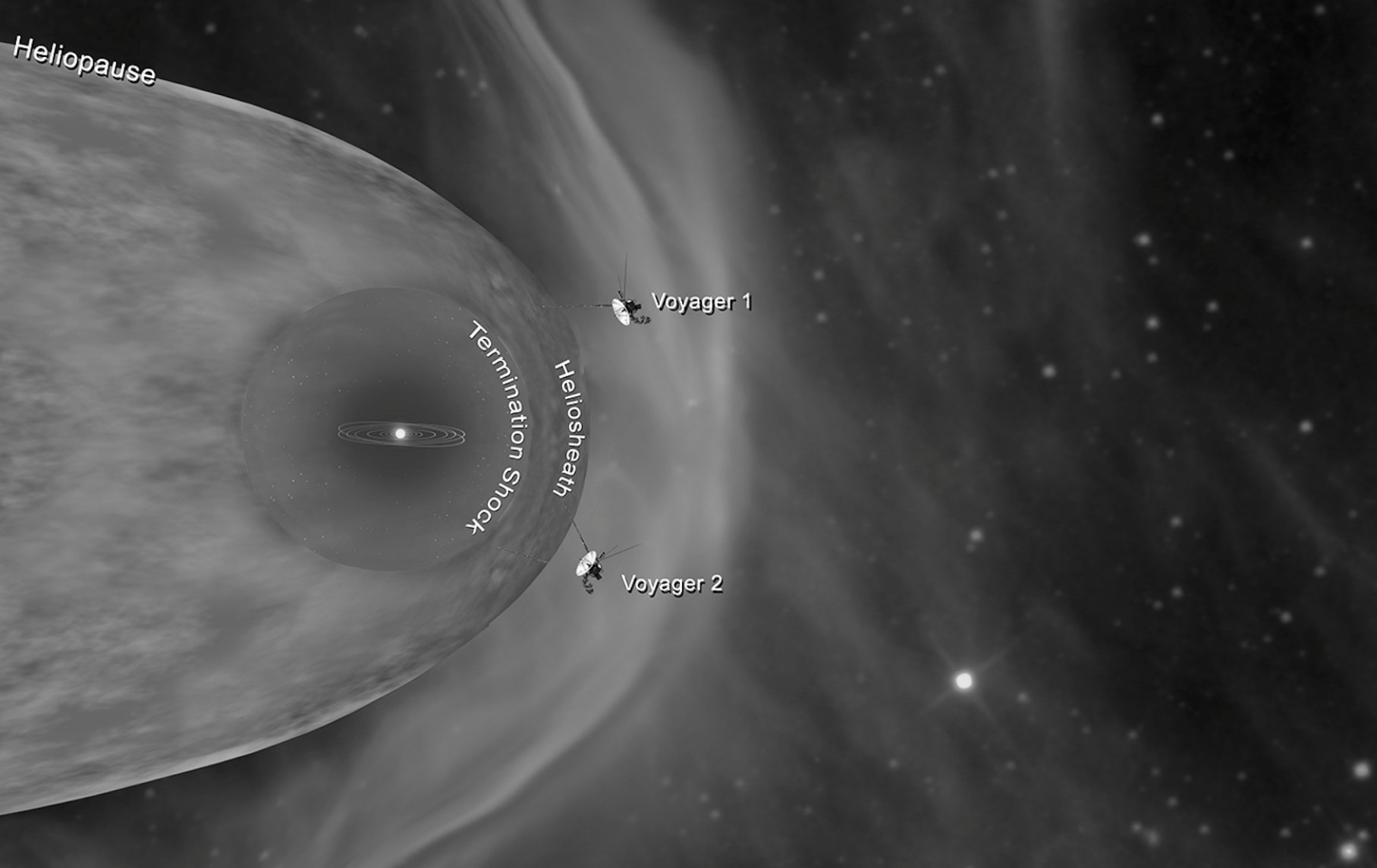}
 \caption{Adopted from Wikipedia, https://en.wikipedia.org/wiki/Heliosphere, and modified to illustrate Voyagers 1 and 2 positions in the LISM in relation to potential extrasolar sources.}\label{VoyagersLISM}
\end{figure}


The energy absorption of electromagnetic waves is highly dependent on the collisionality of the plasma \cite{2019AIPA....9k5205W}. The relatively low intensity of the POE could be due to the conversion efficiency of very low frequency co-production (VLF) to POE. It would also be affected by the absorption of the VLF in accretions near the source, which could significantly increase $r_0$ in \eqref{strainlum}, reducing the VLF intensity. 
The distance from the source determines the geometric dilution and propagation time of the incident electromagnetic wave, but it does not by itself determine the decay time of a plasma oscillation after that wave excites the local LISM.  The persistence of the observed POE must instead be controlled by the duration or recurrence of the driving radiation together with the excitation, damping, and possible continued feeding of plasma modes in the local medium.  If the energy density of a freely damped mode is written as
\begin{equation}
 W(t)=W_{0}\exp(-2\gamma t),
\end{equation}
then its energy e-folding time is $\tau_{E}=\frac{1}{2\gamma}$. The factor of two appears because the wave energy density is proportional
to the square of the oscillation amplitude.  An amplitude decaying as
$e^{-\gamma t}$ therefore corresponds to an energy decay
$W\propto e^{-2\gamma t}$. Persistence over approximately $10$ yr would require an effective damping rate of order
\begin{equation}
 \gamma \sim
  2\times10^{-9}\ {\rm s^{-1}} .
\end{equation}
The studies cited in Refs.~\cite{2024Ap&SS.369...88A,2005A&A...436....9S} identify mechanisms that can produce asymptotically persistent modes, continued energy exchange with background flows, or weak damping. They do not, however, establish this particular damping rate for the plasma sampled by Voyager.  Determining whether the required timescale is realized in the LISM will require a plasma model constrained by the local electron density, temperature, magnetic field, collision rates, and velocity-shear structure.

Without a better understanding of the local properties of the interstellar medium, it is difficult to model the relative contribution to POE persistence and frequency variations from extrasolar radio emissions, plasma oscillations created by these emissions, and local responses to the plasma oscillations in the LISM. It is possible to make some determinations of physical processes in the LISM plasma e.g., \cite{2021AJ....161...11G} and \cite{2005A&A...436....9S}. However, there is no real substitute for in situ observations made by Voyager, and there may not be a comparable mission for many years.

The energies associated with co-production discussed in Section~\ref{NSsource} further complicate the potential contributions of accretion to the POE persistence observed by Voyager. While the co-production intensity is proportional to $h^2$ and generally many orders of magnitude smaller than the GW intensity, near the NS source, the co-production could be substantial. Depending on the local geometry and conversion efficiency, sufficiently intense co-production near the neutron star could perturb or suppress part of the surrounding accretion flow. This could, in theory, provide a signature of co-production that is completely independent of the Voyager observations and is beyond the scope of the present work. A quantitative assessment would require calculations beyond the present plane-wave, order-of-magnitude treatment, potentially including numerical relativity and relativistic magnetohydrodynamics.

A relatively straightforward connection can be made between Voyager cosmic-ray events and expected NS glitches. Voyager detections of cosmic ray events were reported by Ocker et al. \cite{2021NatAs...5..761O} on the order of one per year. This is comparable to, e.g., Vela \cite{Espinoza21} on the order of one every three years and PSR J0922+0638 \cite{Liu25} glitches of about one every 1.5 years. It is possible that there is more than one NS source, which further supports NS as the potential source of Voyager observations. It is also possible that a single NS glitch rate could be comparable to the rate of cosmic-ray events observed by Voyager. Observations of NS glitches are from a population of a few thousand pulsars. This is compared to the approximately 100 million NS in the Galaxy. Only the most energetic fraction of these NS, e.g., the youngest, will produce GW from glitch events \cite{Ball26} at frequencies that could be associated with the Voyager POE detections. Even so, it is not unreasonable to consider the possibility of a single NS source for the Voyager observations from the large population of NS that are not pulsars, e.g., an isolated NS Cassiopeia A \cite{2025ApJ...986..202M} or obscured binary \cite{Pradhan_2026} if the compact object is a NS. Since the Voyager POE observations would be consistent with the highest observed NS spin rates, isolated NS sources would be limited to the most energetic and youngest isolated NS. Identification of Voyager observations of cosmic-ray and POE with extra-solar sources only requires one or perhaps a few nearby and energetic NS.

\section{Neutron star gravitational wave detection} \label{NS GW detection}

The strongest evidence for Voyager POE detections as signatures of co-production would be coincident detections of GW in the kHz range. However, no GW detections have ever been made at frequencies this high, and the source would have to be exceedingly close due to current detector sensitivities at high frequencies. Returning to \eqref{gwFlux_hlum} and \eqref{strainlum} the strain amplitude based on luminosity is,

\begin{equation}
 h = \left( {1.6 \times 10^{-35}~\rm{ \frac{{pc }}{{ W^{\frac{1}{2}}s}}}} \right) \frac{ \mathcal{L}^{\frac{1}{2}} }{ r f}    ~.
\label{strain}
\end{equation}

\noindent Assuming a frequency $f = 1500 ~ \rm{s^{-1}}$, distance of $10 ~ \rm{kpc}$, minimum luminosity of $10^{39} ~ \rm{W}$, and maximum $10^{43} ~ \rm{W}$, the strain amplitude at Earth is on the order of $10^{-23}$ to $10^{-21}$. At a distance of $1 ~ \rm{kpc}$ the strain amplitude at Earth is on the order of  $10^{-22}$ to $10^{-20}$. The nearest pulsars to Earth are about $100 ~ \rm{pc}$ \cite{2004IAUS..218..105L}. At a distance of $100 ~ \rm{pc}$ the strain amplitude at Earth is on the order of  $ 10^{-21}$ to $ 10^{-19}$.  The absence of a reported detection is consistent with the source being beyond current high-frequency sensitivity limits, but does not by itself provide a robust distance constraint. Given LIGO's O2 run's failure to report detections of neutron-star gravitational waves, any potential source should be assumed to be beyond the detector's sensitivity range. We can estimate the minimum range \cite{2018ApJ...857...38H} for excluding potential sources by assuming the detectable characteristic strain would have to be greater than $h_0 \sim 10^{-20}$. To make a rough estimate of the limit on the minimum distance, we use $D \sim 62 ~ {\rm kpc} \left(\frac{h_0}{3.5 \times 10^{-26}} \right)^{-1} \approx 0.2 ~ {\rm pc}$. This extremely low limit on the distance of detectability does not provide any meaningful limitation on a potential neutron star source.

The two distinct observations by Voyager of POE and cosmic-ray events would help narrow both the expected time window for GW observations and the expected frequencies of strong GW production. 
The POE observation could provide a relatively narrow time window for a targeted GW search if the electromagnetic excitation is associated directly with the same transient event.  The cosmic-ray arrival time is less direct because charged particles can experience magnetic deflection and energy-dependent propagation delays.  We therefore use the Voyager cosmic-ray event only as an approximate temporal indicator and do not assume that the cosmic rays and GW must arrive within a fixed $\pm1$-day interval.
Glitch-triggered excitation of a time-dependent quadrupole is one possible mechanism for producing a detectable transient GW signal \cite{Ball26,2024CQGra..41d3001G}.
Identification of potential coincident Voyager detections would greatly narrow the LIGO search window and enable targeted searches for GW signals from NS sources. If the POE are signatures of co-production, the GW would be from highly energetic NS events, e.g., glitches, and could explain the associated POE and cosmic ray detections by Voyager. 

Based on the current literature \cite{2021NatAs...5..761O}, the second LIGO detection run overlaps with Voyager cosmic ray detections in the Summer of 2017. Our examination of the publicly available Voyager database\footnote{\url{https://voyager.gsfc.nasa.gov/rates.html}} places the nominal search window at the beginning of the second week in July 2017 for concurrent GW detections. The lower limit is more difficult to determine because the specific mechanisms of neutron star production of cosmic rays are not completely understood. Further analysis of the Voyager and LIGO databases could identify additional periods of concurrent operation of the LIGO observatories.

The detection of a GW signal near $1.5~\mathrm{kHz}$ coincident with
Voyager POE and cosmic-ray detections would provide strong support for a neutron-star origin of the Voyager POE observations and for the co-production interpretation. Our analysis focused on the Summer of 2017 and LIGO O2 based on the published work of the Voyager collaboration \cite{2021NatAs...5..761O} and the publicly available data analyses tools. This does not consider possible coincident detections from LIGO O3 and O4 runs \cite{2025arXiv251215672M,2026arXiv260314168T,2026arXiv260325808T}, with the recent observations from the O4 run \cite{2025arXiv250818079T}. A more advanced joint analysis of the Voyager and LIGO observations is required to identify any statistically significant coincidence.

\section{Discussion}

Very early in the Voyager mission, Kurth et al.~\cite{Kurth1984} speculated that the source of the $3~\mathrm{kHz}$ radio emissions detected by Voyager in the 1980s might be an extrasolar neutron star, but no acceptable theory existed at the time to explain the observations. In our earlier work \cite{PhysRevD.96.124030,10.1142/S0218271818470211} we demonstrated that GW from the central compact objects in core-collapse supernovae should leave detectable signatures of co-production in the structure and evolution of the supernova remnants. This recently developed theory of co-production provides a possible extrasolar mechanism for generating radiation at frequencies relevant to the plasma oscillations detected by Voyager.
Co-production provides a possible mechanism consistent with the observed POE frequencies and intensities under reasonable assumptions about coupling to the local plasma \cite{10.1142/S0218271818470211,2010MNRAS.406..863L,2022SSRv..218...27M}. 

It is important to distinguish the semiclassical co-production mechanism considered here from the perturbative quantum process $g+g\rightarrow\gamma+\gamma$.  The latter is characterized by an extremely small scattering cross section, whereas the former is obtained by solving the covariant Maxwell equations in a prescribed gravitational-wave background.  In the semiclassical calculation, the relevant measure is therefore not a graviton-to-photon scattering probability, but the ratio of the electromagnetic and gravitational-wave energy fluxes.  For the weak, plane-wave background considered in Section~\ref{NSsource}, this ratio is $\frac{F_{\rm em}}{F_{\rm gw}}=2h^2$, and is consequently small when $h\ll 1$.  The leading electromagnetic term in \eqref{OutState} has twice the frequency of the gravitational-wave background.  Thus, an electromagnetic signal near $3~{\rm kHz}$ is theoretically consistent with a gravitational-wave source near $1.5~{\rm kHz}$.  This frequency relation is a prediction of the semiclassical model, however, and does not by itself establish that the Voyager signal was produced by gravitational waves.

Voyager's cosmic ray detections, preceding the POE detections, are consistent with co-production from energetic NS gravitational waves. The potential identification of NS as the source of cosmic ray detections is beyond the scope of the current work but would support NS as the source of POE observations. The strongest possible confirmation of co-production as the source of radio emissions detected by Voyager, in the LISM, would be coincident detections of GW. But our focus here is on the identification of Voyager detections of POE as possible signatures of co-production. The data analysis required to identify POE and GW coincidence observations is intriguing but beyond the scope of this work. Even in the absence of confirmed coincident GW observations, the estimates presented here show that co-production remains a possible source of the radio emission detected by Voyager.

As discussed in Section~\ref{NSsources}, sufficiently intense co-production could perturb or suppress part of the surrounding accretion flow, depending on the local geometry and conversion efficiency. This could limit the duration of any accretion-supported continuous GW emission. The observed POE persistence could nevertheless be governed primarily by the response of the rarefied LISM plasma.

Identification of signatures of co-production in the LISM might require future science missions to conclusively observe coincident GW and LISM plasma oscillations. In particular, GW observatories with higher sensitivities \cite{2019PhRvD..99j2004M} in the kHz range and space missions to the LISM, including plasma-wave detectors. Right now, no such missions are in development, but if they do occur, they could make important contributions to our understanding of fundamental physics, the Solar system, and our Galactic neighborhood.

\section*{Acknowledgment} The authors would like to thank William Kurth and Alen Cummings for help with acquiring and interpreting the Voyager online database. PJ would like to thank Logan Finke, Andri Gretarsson, Joseph Ribaudo, and Ryan Totman for helpful discussions.

\section*{Data availability}

All data used for the calculations in this paper are publicly available from the Voyager Scientific Mission at \cite{Kurth_2023} and https://voyager.gsfc.nasa.gov/rates.html.

\section*{Funding} DS received support from the Frank Sutton Research Fund and a California State University Fresno Research, Scholarship, and Creative Activity Award.

\section*{} Corresponding author: Preston Jones \\
Email: jonesp13@erau.edu


\bibliography{CCSNe}

\end{document}